\documentclass{PoS}
\usepackage{epsf}

\title{
\vspace*{-2cm}
\begin{flushright}\it \normalsize IPPP/10/05 \end{flushright} \vspace{2cm}
Resonant particle production at hadron colliders}

\ShortTitle{Resonant particle production}

\author{Pietro Falgari, Paul Mellor and \speaker{Adrian Signer}\\
        Institute for Particle Physics Phenomenology\\
        Durham University\\
        United Kingdom\\
        E-mail: \email{pietro.falgari@durham.ac.uk, \\
                       paul.mellor@durham.ac.uk, \\
                       adrian.signer@durham.ac.uk}}

\abstract{We present a method to compute off-shell effects for
  processes involving resonant particles at hadron colliders with the
  possibility to include realistic cuts on the decay products. The
  method is based on an effective theory approach to unstable particle
  production and, as an example, is applied to $t$-channel single top
  production at the LHC.}

\FullConference{RADCOR 2009 - 9th International Symposium on 
Radiative Corrections (Applications of Quantum Field 
Theory to Phenomenology) ,\\
		 October 25 - 30 2009\\
		 Ascona, Switzerland}

\begin{document}

\section{Introduction}

Heavy unstable particles such as the top quark, $Z$ and $W$ bosons as
well as virtually all hypothetical particles in
beyond-the-Standard-Model scenarios can only be studied through their
decay products. The simplest way to compute cross sections involving
such particles is to first consider on-shell production of the heavy
particle and then its subsequent decay. Taking into account the matrix
element of the decay of the on-shell particle --- this is sometimes
called the improved narrow-width approximation --- it is possible to
apply realistic cuts on the decay products. In this framework we can
go to higher orders in perturbation theory by computing separately
corrections to the production and the decay of the on-shell heavy
particle. While this approximation is very often good enough, this
talk addresses the question on how to go beyond and include off-shell
effects in a systematic way. Thus we are led to consider processes with
resonant (nearly on-shell) rather than on-shell heavy particles.

Even though the framework for computing off-shell effects presented
here is quite general, let us consider an explicit example,
$t$-channel single top production. More precisely we consider the
partonic process $u(p_1)\, b(p_2)\to d(p_3) b(p_4)\, W^+(p_W)\to
d(p_3)\, b(p_4)\, e^+(p_5)\, \nu(p_6)$, where the invariant mass of
the top decay products is understood to be close (but not necessarily
equal) to the top mass, i.e. $(p_4+p_W)^2 = m_t^2 + \Delta$ with
$\Delta/m^2_t \ll 1$. The decay $W^+\to e^+\, \nu$ is taken into
account in the improved narrow-width approximation. It is clear that
the dominant contribution to this process with its kinematic
constraint is given by production and subsequent decay of a top
quark. Self-energy resummation avoids the non-integrable singularity
in the top propagator at $(p_4+p_W)^2 = m_t^2$ by changing the
denominator of the propagator to $(p_4+p_W)^2 - m_t^2+i m_t\Gamma_t$,
where $\Gamma_t$ is the width of the top. But even at tree-level, at
some point we will have to take into account so called background
diagrams which do not refer to a top quark at all. The situation
becomes even more involved if we want to include higher-order
effects. This can be done using the pole
approximation~\cite{Aeppli:1993rs}, which can be considered as a first
step towards a systematic expansion of the amplitude in $\Gamma_t/m_t$
around the complex pole. Within the pole approximation the one-loop
corrections can be split in a gauge-invariant way into factorizable
and non-factorizable corrections~\cite{Fadin:1993dz}. Factorizable
corrections correspond to corrections to the production and decay part
of the unstable particle, whereas non-factorizable corrections link
the two parts. Even though non-factorizable corrections (and off-shell
effects in general) have been studied extensively in the
literature~\cite{nf-calc} they are usually neglected for processes at
hadron colliders, because in most cases they are found to be
small~\cite{nf-small}. In fact there are large
cancellations~\cite{nf-theorems} that are partly responsible for the
smallness of the corrections. However, with cuts in the final state
these cancellations are not perfect any longer and, as stated above,
it is the purpose of this work to study these corrections.

\section{Effective theory and virtual corrections \label{virtual}}

The salient feature in the problem at hand is the presence of
different scales, $m_t \gg \Gamma_t$. This calls for using an
effective theory (ET) approach to the problem. Indeed, it has been
found~\cite{Chapovsky:2001zt} that in an ET approach factorizable
corrections simply correspond to the hard corrections. In this
context, hard is understood as a mode using the method of
regions~\cite{Beneke:1998zp} and means momenta that scale as $p\sim
m_t$. The standard procedure within an ET approach is to first
integrate out the hard modes and thereby obtain an effective
Lagrangian. This Lagrangian consists of gauge invariant operators
multiplied by matching (Wilson) coefficients, which have a
perturbative expansion in the coupling and are gauge independent as
well. The matching is done on-shell with $p_t^2 = m_t^2- i\, m_t
\Gamma_t$. While the matching coefficients take into account the
factorizable corrections, the non-factorizable corrections are
reproduced in the effective theory by the still dynamical soft (and
possibly other) modes. A soft mode corresponds to a momentum scaling as
$p\sim m_t \delta$ where we use $\delta \sim \Delta/m^2_t \sim
\Gamma_t/m_t \sim \alpha_{\rm ew} \sim \alpha^2_{\rm s}$ to
generically denote a small quantity. The construction of an effective
Lagrangian is a standard procedure and has been used in many different
contexts. For the application to unstable particles it has been
explained in detail in Ref.~\cite{Beneke:2004km}. We thus restrict
ourselves to a few remarks relevant for our example.

\begin{figure}[t]
   \epsfysize=3.2cm
   \centerline{\epsffile{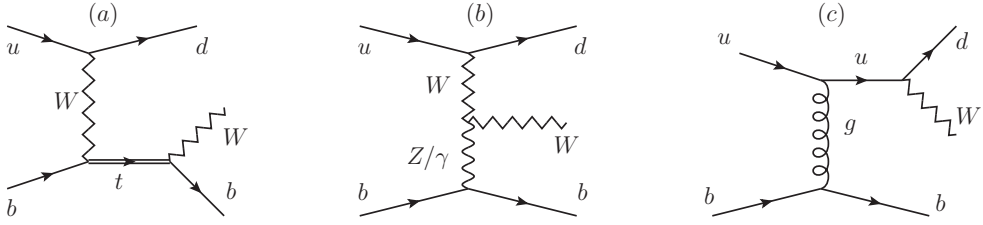}}
   \caption{(a) resonant tree-level diagram contributing to
     $A^{(3,0)}_{(-1)}$ and $A^{(3,0)}_{(0)}$; (b) example of a
     background diagram contributing to $A^{(3,0)}_{(0)}$; (c) example
     of a QCD diagram contributing to $A^{(1,2)}$. \label{Fig0}}
\end{figure}

The (strictly fixed-order) tree-level amplitude is a function of
$m_t$, other masses and the momenta $p_1\ldots p_6$. Choosing $\Delta
\equiv (p_W+p_b)^2-m_t^2$ as one of the independent kinematic
variables and expanding around the pole in $\Delta$ we write
\begin{equation}
{\cal A}^{{\rm tree}} = \delta_{31} \delta_{42} \Big(  
g_{ew}^3\, A^{(3,0)}_{(-1)} + g_{ew}^3\, A^{(3,0)}_{(0)} +
\ldots \Big) +
T^a_{31} T^a_{42}\,g_{ew} g_s^2\, A^{(1,2)}
\label{treeA}
\end{equation}
The leading part, $g_{ew}^3\,A^{(3,0)}_{(-1)}$, receives contributions
solely from the resonant diagram, part~(a) of Figure~\ref{Fig0}, and
scales as $\delta^{1/2}$, whereas for the subleading part $g_{ew}^3\,
A^{(3,0)}_{(0)} \sim \delta^{3/2}$ also background diagrams have to be
taken into account. Given that ${\cal A}^{{\rm tree}}$ is gauge
invariant it is clear that each term in the expansion in
Eq.~(\ref{treeA}) is separately gauge invariant.  Note that we also
include the QCD contribution $g_{ew} g_s^2\, A^{(1,2)} \sim \delta$
which is usually considered to be a background. However, from our
point of view this part has the same final state and contributes to
${\cal A}$, albeit only at subleading order. Squaring the amplitude we
obtain
\begin{equation}
M^{{\rm tree}} = 
g_{ew}^6\, N^2_c\,
  \left|A^{(3,0)}_{(-1)}\right|^2
+ g_{ew}^6\, N^2_c\,   
2\, {\rm Re}\left(A^{(3,0)}_{(-1)}\,[A^{(3,0)}_{(0)}]^*\right)
+ g_{ew}^2 g_s^4 \, N_c\, C_F/2 
\left|A^{(1,2)}\right|^2 + \ldots
\label{treeM}
\end{equation}
This is simply an expansion of $M^{{\rm tree}}$ in the kinematic
variable $\Delta$ which we constrain to be small. The (first) leading
term scales as $g_{ew}^6\, \Delta^{-2} \sim \delta$. The subleading
electroweak and the QCD term scale as $g_{ew}^6\, \Delta^{-1} \sim
\delta^2$ and $g_{ew}^2 g_s^4 \sim \delta^2$ respectively, and
omitted terms are further suppressed in $\delta$. This expansion is
not complete yet, because it takes into account only the small
parameter $\Delta/m_t^2 \sim \delta$ but not $\alpha_{\rm ew} \sim
\Gamma_t/m_t \sim \delta$. The two small parameters are intrinsically
linked and when the corresponding expansions are combined the ET
becomes a very powerful tool.

%%%%%%%%%%%%%%%%%%%%%%
%%%%%%%%%%%%%%%%%%%%%%

Within the ET framework, the leading contribution to ${\cal A}^{{\rm
    tree}}$ is reproduced in three steps: production of an on-shell
top (through an operator in the ET with its matching coefficient),
propagation of a resonant top (through the leading bilinear top
operator) and decay of an on-shell top ( through a decay operator in
the ET with its matching coefficient). The bilinear operator resums
the (gauge-invariant) hard part of the self energy. The subleading
terms give rise to 5-point operators. Again, the operators as well as
their matching coefficients are separately gauge invariant.

\begin{figure}
   \epsfysize=1.8cm
   \epsffile{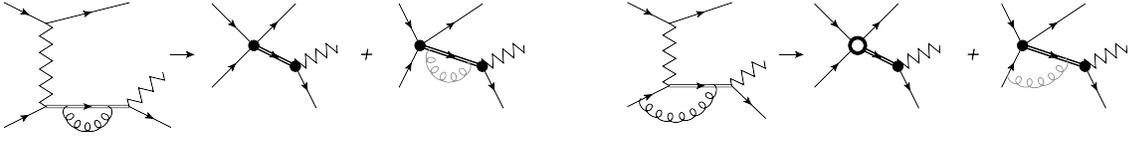}
   \caption{Two examples of how usual loop diagrams are reproduced by
     the ET. Grey lines indicate soft gluons, solid (hollow) dots
     indicate tree-level (one-loop) matching
     coefficients. \label{FigLoop}}
\end{figure}

We aim at computing the cross section up to ${\cal O}(\delta^{3/2})
\sim {\cal O}(\alpha_{\rm s} \delta)$, which corresponds to including
one-loop QCD corrections to the leading resonant part. Within the ET
approach this will amount to computing one-loop corrections to the
Wilson coefficients for the leading production and decay operators, as
well as explicit calculation of one-loop corrections within the
effective theory. To illustrate the connection between the ET and
standard loop calculations, let us consider two examples, shown in
Figure~\ref{FigLoop}.

%%%%%%%%%%%%%%%%%%%%%%
%%%%%%%%%%%%%%%%%%%%%%
Example (a) illustrates how the self-energy is split into a hard part
and a soft part. The hard part scales as ${\cal O}(\delta^0)$ and,
therefore, has to be resummed. The soft part is suppressed by an
additional power $\delta^{1/2}$ with respect to ${\cal A}^{{\rm
    tree}}$. Thus we need to include only one soft self-energy
insertion at ${\cal O}(\delta^{3/2})$.  Example (b) illustrates how a
normal loop diagram is split into a hard (factorizable) contribution
included in the matching coefficient and a soft (non-factorizable)
contribution where the gluon is still dynamical in the ET. The
situation is not affected by including an arbitrary number of hard
self-energy contributions to the top propagator, but an additional
soft self-energy insertion results in a contribution beyond ${\cal
  O}(\delta^{3/2})$.

The ET allows us to identify and compute the minimal set of
corrections needed at a certain order in $\delta$. In particular a
full one-loop calculation including all non-resonant diagrams is not
needed at this order. The price to be paid for this simplification is
that we have to insist on the kinematical constraint $\Delta/m_t^2 \ll
1$. A corresponding calculation using e.g. the complex mass
scheme~\cite{Denner:2006ic} would be more complicated, but would allow
us to study observables without this constraint.

\section{Real corrections}

Using an ET approach as described in the previous section is a well
established tool for the calculation of total cross
sections~\cite{totXS}.  When computing real corrections for more
general observables within an ET approach we run into
difficulties. The problem is related to the fact that an ET approach
relies on knowing and making explicit all scales of the problem. As
long as we have not precisely defined the observable we are interested
in, this is not the case. 

In Section~\ref{virtual} we based the expansion on the small scale
$\Delta$ (and the couplings), implicitly assuming there are no further
small scales. While this does put some restrictions on possible
observables (similar to any fixed-order calculation) it allows us to
study a large class of physical quantities.
In the case of real corrections, we have an additional gluon present
in the final state. Depending on $p_7$, the momentum of the gluon, the
expansion parameter does change. Consider e.g. the case where $p_7$ is
collinear to $p_4$, the momentum of the outgoing $b$~quark. In this
case $p_7$ and $p_4$ will combine to a jet and the proper expansion
parameter would be $(p_4+p_7+p_W)^2 - m_t^2$. On the other hand, if
$p_7$ is well resolved, this will give rise to an additional (gluon)
jet and the proper expansion parameter would be $(p_4+p_W)^2 - m_t^2$,
as in Section~\ref{virtual}.

In order to deal with this we deviate from a strict ET approach. Using
the subtraction method to compute the real corrections we write
\begin{eqnarray}
\int d\Phi_{n+1}  M_{n+1} &=& 
\int d\Phi_{n+1} \left(M_{n+1}- M^{\rm sing}_{n(+1)} \right)
+ \int d\Phi_{n+1}\,  M^{\rm sing }_{n(+1)}
\label{realM}\\
&\simeq& 
\int d\Phi_{n+1} \left( M_{n+1} - M^{\rm sing}_{n(+1)} \right)
+ \int d\Phi_{n+1}\,  M^{\rm sing\ {exp} }_{n(+1)}
\nonumber
\end{eqnarray}
Here $M_{n+1}$ is the matrix element squared for the process $u\, b
\to d\, b\, e^+\, \nu\, g$ and $M^{\rm sing}_{n(+1)}$ denote the
limits of the matrix elements in the singular (soft and collinear)
regions. Upon phase-space integration the term $\int d\Phi_{n+1}
M^{\rm sing }_{n(+1)}$ would match the infrared singularities of the
full virtual one-loop amplitude. Since we use only the expanded
virtual term, there is a mismatch between the singularities.
Fortunately, the singular limits have only $n$-parton kinematics and
we can use the same expansion as for the tree-level and virtual
amplitudes. Computing $\int d\Phi_{n+1}\, M^{\rm sing\ {exp}
}_{n(+1)}$, i.e. integrating the expanded singular limits over the
phase space, produces the same infrared singularities as the expanded
virtual corrections. In the first term on the r.h.s. of
Eq.~(\ref{realM}) we do not perform any expansion. The mismatch
between what we subtract and what we add back is subleading in
$\delta$ and beyond the accuracy we are aiming at.

In principle we should use the full matrix element $M_{n+1}$ in
Eq.~(\ref{realM}) to ensure gauge invariance. We can simplify the
calculation by taking only resonant real diagrams. This potentially
introduces a gauge dependence which, however, is beyond the accuracy
of our calculation. Thus the situation is completely analogous to the
renormalization or factorization scale dependence, which is widely
accepted as long as it is beyond the accuracy of the calculation. A
variation of $\mu$ within a reasonable window tells us as much (or as
little) about the size of neglected higher-order terms as a variation
of the gauge parameter $\xi$ within a window around 1. Of course, it is
always possible that there are terms that are parametrically of higher
order, but numerically important. This is usually associated with the
presence of widely different scales and requires additional
resummations.

\section{Results and outlook}

We are now ready to compute an arbitrary infrared-safe quantity at
${\cal O}(\delta^{3/2})$. As always, the word arbitrary has to be
taken with some caution. First of all we have to enforce the
constraint $\Delta \sim m_t \Gamma_t$. Second, we implicitly assume
there are no other small scales introduced through the observable.  A
similar requirement is present for any fixed-order calculation.

As an example we consider $u\, b \to W$ plus jets, defining the jets
using a $k_\perp$ cluster algorithm. We require a $b$~jet, $J_b$, and
at least one more jet, both with $p_\perp > 20$~GeV. In addition we
require $150~{\rm GeV} \le M^{\rm inv} \le 200$~GeV, where $M^{\rm
  inv} =\sqrt{(p_W+p_{J_b})^2}$ is the invariant mass of the ($W\, J_b$)
pair. We stress that we could add cuts on the decay products of the
$W$ or any other 'reasonable' cuts.

\begin{figure}[t]
   \centerline{
   \epsfysize=4.3cm
   \epsffile{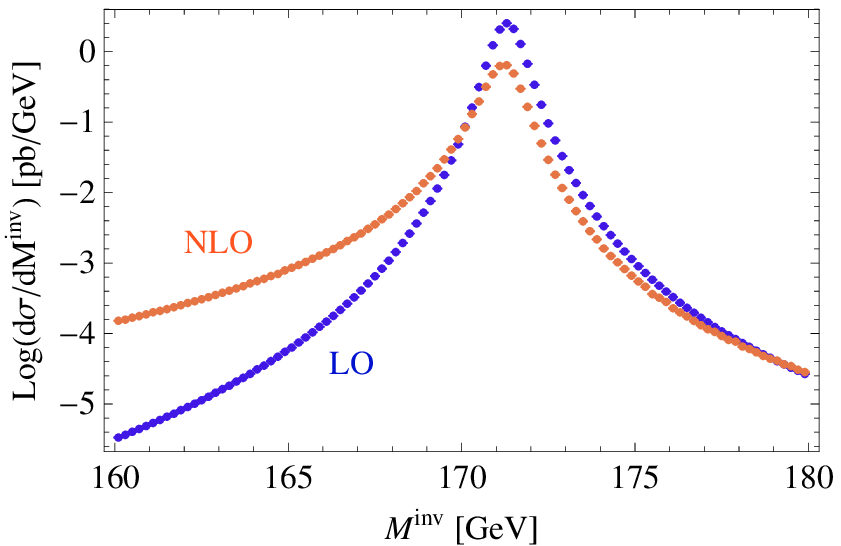}
   \qquad
   \epsfysize=4.3cm
   \epsffile{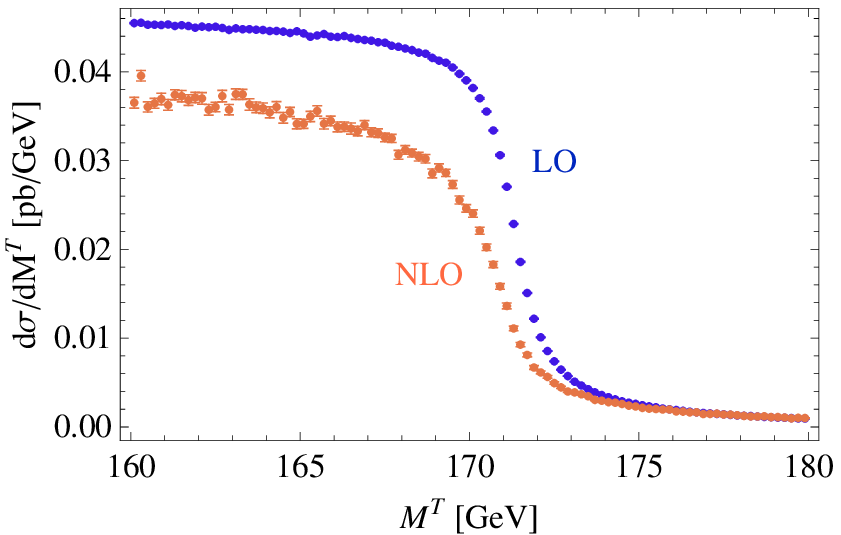}}
 \caption{Invariant and transverse mass distribution at ${\cal
     O}(\delta)$ (blue, dark) and ${\cal O}(\delta^{3/2})$ (red,
   light). \label{FigPlot}}
\end{figure}

In Figure~\ref{FigPlot} we show the $M^{\rm inv}$ and $M^T$
distribution for the LHC with $\sqrt{s} = 10$~GeV, where $M^T$ is the
transverse mass of the ($W\, J_b$) system. We set $\mu_F=\mu_R =
100$~GeV, $m_t=171.3$~GeV and $\Gamma_t = 1.32$~GeV and use NLO MSTW
pdfs~\cite{Martin:2009iq} for the LO and NLO results. Compared to
previous NLO calculations~\cite{singletop} our result also includes
non-factorizable corrections which leads to a small but visible
deviation from a Breit-Wigner shape in the in the $M^{\rm inv}$
distribution.

%%%%%%%%%%%%%%%%%%%%%%
%%%%%%%%%%%%%%%%%%%%%%

Of course, the results shown in Figure~\ref{FigPlot} are not complete,
since at NLO we also have to include gluon-initiated processes and
there are other partonic initial states such as $b\, u\to d\, b\, W$
and $c\, b\to s\, b\, W$. A full analysis as well as the potential
impact of non-factorizable corrections on a measurement of $m_t$ for
single top and in particular $t\bar{t}$ pair production is left for
future work.

\end{document}